\documentclass[iop]{emulateapj}

\usepackage{amsfonts}
\usepackage{amsmath}
\usepackage{natbib}

\renewcommand{\[}{\begin{equation}}
\renewcommand{\]}{\end{equation}}

\let\boldgrk=\gkvecten
\let\boldgrksc=\gkvecseven

\def\gkthing#1{{\mathchoice%
	{\hbox{{\boldgrk\char#1}}}
	{\hbox{{\boldgrk\char#1}}}
	{\hbox{{\boldgrksc\char#1}}}
	{\hbox{{\boldgrksc\char#1}}}}}

\def\vtheta{\gkthing{18}}

\def\vOmega{{\bf\Omega}}
\def\vDelta{{\bf\Delta}}
\def\rs{{\rm s}}
\def\i{{\rm i}}
\def\si#1{\psi^{(#1)}}\def\ho#1{\rho^{(#1)}}\def\Igma#1{\Sigma^{(#1)}}
\def\Rg{R_{\rm g}}
\def\bm#1{{\bf#1}}
{\newif\ifnotend
\notendtrue
\def\veclist{ABCDEFGHIJKLMNOPQRSTUVWXYZabcdefghijklmnopqrstuvwxyz.}
\def\top#1#2.{#1}
\def\tail#1#2.{#2.}
\loop\expandafter\xdef\csname v\expandafter\top\veclist\endcsname%
{{\noexpand\bf\expandafter\top\veclist}}
\edef\veclist{\expandafter\tail\veclist}
\if\veclist.\notendfalse\fi\ifnotend\repeat}

\def\fracj#1#2{{\textstyle{#1\over#2}}}
\def\pa{\partial}

\def\e{{\rm e}}
\def\d{{\rm d}}

\shorttitle{Orbital diffusion in stellar discs}
\shortauthors{Fouvry et al.}

\begin{document}

\setcounter{tocdepth}{3}

\title{Self-gravity, resonances and orbital diffusion in stellar discs}

\author{Jean-Baptiste Fouvry\altaffilmark{1,2}, James Binney\altaffilmark{3} and Christophe Pichon\altaffilmark{1,2}}

\altaffiltext{1}{Institut d'Astrophysique de Paris, CNRS (UMR-7095), 98 bis Boulevard Arago, 75014, Paris, France}
\altaffiltext{2}{UPMC Univ. Paris 06, UMR-7095, 98 bis Boulevard Arago, 75014, Paris, France}
\altaffiltext{3}{Rudolf Peierls Centre for Theoretical Physics, University of Oxford, Keble Road, Oxford OX1 3RH}

\begin{abstract}
Fluctuations in a stellar system's gravitational field cause the orbits of
stars to evolve. The resulting evolution of the system can be computed with
the orbit-averaged Fokker-Planck equation once the diffusion tensor is known.
We present the formalism that enables one to compute the diffusion tensor
from a given source of noise in the gravitational field when the system's
dynamical response to that noise is included. In the case of a cool stellar
disc we are able to reduce the computation of the diffusion tensor to a
one-dimensional integral. We implement this formula for a tapered Mestel disc
that is exposed to shot noise and find that we are able to explain
analytically the principal features of a numerical simulation of such a disc.
In particular the formation of narrow ridges of enhanced density in action
space is recovered.  As the disc's value of Toomre's $Q$ is reduced and the
disc becomes more responsive, there is a transition from a regime of heating
in the inner regions of the disc through the inner Lindblad resonance to one
of radial migration of near-circular orbits via the corotation resonance in
the intermediate regions of the disc.  The formalism developed here provides
the ideal framework in which to study the long-term evolution of all kinds of
stellar discs. 
\end{abstract}


\section{Introduction}
\label{sec:introduction}

Many, perhaps all, stars are born in a stellar disc. Major mergers destroyed
some discs quite early in the history of the Universe, but many others have
survived to the present day, including the disc of which the Sun is a part. Hence
an understanding of the dynamics and evolution of stellar discs is an
essential ingredient of cosmology. Conversely, cosmology provides the
framework within which disc dynamics should be studied because dark-matter
halos make large contributions to the gravitational fields in which discs
move, and dark-matter substructures are major contributors to the
gravitational noise to which discs are exposed.

Serious study of disc dynamics got underway in the 1960s with seminal works by
Lin, Shu, Goldreich, Toomre and Lynden-Bell.  Although some important insights
were gained at that stage, fundamental questions were left open. While the
earliest work was almost entirely analytic in nature, numerical simulations
of stellar discs became more important over time, and revealed important
aspects of disc dynamics that were hard to understand analytically. In
particular it emerged that discs that are completely stable at a linear level
nevertheless  develop spiral structure that eventually grows to amplitudes of
order unity, so the disc becomes something like a barred spiral galaxy
\citep[][hereafter S12]{Sellwood2012}.
\cite{SellwoodCarlberg14} recently offered a convincing explanation of this
phenomenon that hinges on the fact that resonances localise the impact that
a fluctuation has on a disc. This localisation is the major focus of this
paper.

Self-gravitating discs are responsive dynamical systems, in which (a)
rotation provides an abundant supply of free energy, and (b) resonances play
a key role. The ready availability of free energy leads to some stimuli being
powerfully amplified, while resonances localise the dissipation of free
energy with the result that even a very small stimulus can result in a disc
evolving to an equilibrium that is materially different from the one from
which it started. 

The stimuli to which discs respond are various sources of gravitational
noise. They include, Poisson noise arising from the finite number of stars in
a disc, Poisson noise arising from the finite number of giant molecular
clouds in the interstellar medium, and Poisson noise arising from the finite
number of massive sub-halos around a galaxy. Spiral arms in the distribution
of gas provide another source of gravitational noise, while the rotating
gravitational field of a central bar constitutes a source of stimulus that is
more systematic than noisy. The history of a real stellar disc will largely comprise
responses to all these stimuli. 

In the solar neighbourhood at least three
distinct manifestations of such responses are evident:

\begin{itemize}
\item[(i)] The random velocity of each
coeval cohort of stars increases with the cohort's age \citep{Wielen77,AumerB}.

\item[(ii)] The velocity
distribution at the Sun contains several ``streams'' of stars
\citep{Dehnen98}. Each such
stream contains stars of various ages and chemistries that are all 
responding to some stimulus in a  similar way \citep{Famaey05}.

\item[(iii)] In the two-dimensional space in which one coordinate is angular
momentum $J_\phi$ and the other is a measure of a star's radial
excursions, such as the radial action $J_r$,  the density of stars shows
elongated features.  The density of stars is depressed near $J_r=0$
but enhanced at larger $J_r$ in such a way that the whole region of disturbed
stellar density forms a curve that is consistent with being a curve on which
a resonant condition such as $2\Omega_\phi-\Omega_r=\hbox{constant}$ holds
\citep{Sellwood10,McMillan13}. We shall call  a feature of this type a {\bf
resonance ridge}.  \cite{SellwoodCarlberg14} have argued that resonance
ridges play a crucial role in the long-term dynamics of stellar discs.

\end{itemize}

Numerical simulations of stellar discs are extremely challenging because the
near two-dimensional geometry of discs combined with their responsive nature
causes discreteness noise to be dynamically important unless the number of
particles employed exceeds $\sim200\,000$. Hence only recently has it become
straightforward to simulate a disc with a sufficient number of particles for
discreteness noise to be dynamically unimportant for many dynamical times
(S12).  It is particularly hard to simulate accurately a disc
that is embedded in a cosmological simulation and thus exposed to cosmic
noise. Moreover, the utility of a simulation is greatly increased if one
understands analytically why it evolves the way it does. A goal of this paper
is to show the extent to which perturbation theory explains a phenomenon --
resonance ridges -- that is seen in both numerical simulations and surveys of
the solar neighbourhood.

Perturbation theory is much more than a device for computing approximate
solutions to equations: throughout physics it provides the conceptual
framework we use to {\it understand\/} phenomena. Examples include the
concepts of a free particle and an interaction in particle physics, a phonon
and a gravity wave in condensed-matter physics, semi-major axis and
eccentricity in planetary dynamics, and so on. The natural way to increase
our understanding of the dynamics of stellar discs is to practise the
application of perturbation theory to these systems, so we may gain insight
into how these fascinating systems work, and learn how one can think about
them most profitably.

\cite{Kalnajs} laid the foundations of perturbation theory for stellar discs.
The theory is based on the use of angle-action coordinates -- the coordinates
that were introduced to understand the dynamics of the solar system.  These
coordinates are being increasingly used to build equilibrium models of hot
and cold stellar systems \citep{Binney10,Binney14,Piffl14}, and to study the
dynamics of star streams
\citep{HelmiWhite99,Sellwood10,McMillan13,EyreBinney,SandersBinney}.
\cite{Binney1988} used these coordinates to derive the orbit-averaged
Fokker-Planck equation for a stellar disc. However, they did not consider the
origin of the fluctuations in the gravitational potential that drive stellar
diffusion. \cite{Weinberg2001a} divided the driving fluctuations into the
contribution from some external stimulus, and the self-consistent dynamical
response of the system itself to the stimulus.  Weinberg's treatment was
adapted to systems that are spherical when unperturbed, while here we
restrict ourselves to razor-thin discs, in which case the construction of the
angle-action coordinates is trivial.

The paper is organised as follows. Section~\ref{sec:diffusionequation}
recalls from~\cite{Binney1988} and~\cite{Weinberg2001a} the general
principles of secular evolution, the orbit-averaged Fokker-Planck equation,
and the use of a set of biorthonormal potential-density pairs to compute the
diffusion tensor that is jointly generated by an external stimulus and the
system's response to this stimulus. Section~\ref{sec:appl} specialises this
formalism to a razor-thin, cool disc by introducing a set of basis functions
that comprise localised, tightly-wound spirals.  Using these basis functions
we are able to reduce the computation of the diffusion tensor, which in
principle requires a double sum over basis functions, to a single integral
over radial wavenumbers. In Section~\ref{sec:model} we compute the diffusion
tensor for a tapered Mestel disc that is excited by shot noise, and show that
the resulting predictions for the disc's evolution reproduce the main
features of the N-body simulations reported by S12. Finally, we conclude in
Section~\ref{sec:conclusions}.

\section{Fluctuations and secular evolution}
\label{sec:diffusionequation}

To zeroth order, stellar discs are systems that have achieved statistical
equilibrium within an axisymmetric gravitational field that arises not only
from their mass but also from mass contained in other components of the
galaxy, especially the bulge and the dark halo. The Hamiltonian associated
with the field is to a good approximation integrable, so all orbits may be
assumed to admit three isolating integrals, which we take to be the actions:
$J_\phi=L_z$ is the angular momentum about the field's symmetry axis; $J_r$,
which quantifies the amplitude of a star's radial oscillations; and $J_z$,
which quantifies oscillations perpendicular to the field's equatorial plane
\citep{born1960mechanics,BT08}. On account of the integrability
of the gravitational field and Jeans' theorem,  we can assume that at each
instant the disc's distribution function (DF) is a function $f(\vJ,t)$ of the
actions only, rather than a general function on phase space
$f(\vJ,\vtheta,t)$, which has dependence on the variables that are
canonically conjugate to the actions, namely the angle variables $\theta_i$.

Any fluctuation in the gravitational field causes each star to deviate from
its original orbit $\vJ$ and to settle after the fluctuation has died away on
another orbit $\vJ'=\vJ+\vDelta$. Hence fluctuations cause stars to diffuse
through action space. Since initially action space is populated by stars only
along the $J_\phi$ axis, this diffusion raises the density of stars away from
this axis by populating orbits with distinctly non-zero $J_r$. As a
consequence, the velocity dispersion within the disc rises, so fluctuations
``heat'' the disc. Stars also diffuse along the $J_\phi$ axis.  Since such
diffusion merely transfers stars from one nearly circular orbit to another of
a different radius, this component of diffusion is called radial migration.
Radial migration does not heat the disc, and given that the density of stars
does not vary rapidly along the $J_\phi$ axis, it can easily go unnoticed.
However, chemical evolution within the disc establishes a radial gradient in
metallicity, and radial migration is most readily detected through its
interaction with this gradient \citep{SellwoodB}: radial migration tends to
erase the correlation between the ages and metallicities of stars near the
Sun by bringing to the Sun both old, metal-rich stars formed at small radii
and young metal-poor stars formed at large radii.

Fundamentally fluctuations drive the long term (``secular'') evolution of
discs in much the same way that they drive the much better understood
secular evolution of globular clusters, but resonances are unimportant in
globular clusters and dominant in discs. As indicated in the Introduction,
resonances localise  the impact of fluctuations and give rise to ridges
in action space that are the primary focus of this paper. However,
\cite{SellwoodB} pointed out that at the corotation resonance, stars are
scattered at constant radial action, i.e., parallel to the $J_\phi$ axis. So
scattering at corotation does not give rise to a ridge and is likely to go
overlooked if one does not pay attention to the metallicities of stars. While
at corotation only $J_\phi$ changes, at a Lindblad resonance both $J_r$ and
$J_\phi$ change, so radial migration is not confined to corotation, as is
often stated.

\subsection{Orbit-averaged Fokker-Planck equation}

Since we are imagining that stars are conserved, the equation that governs
the secular evolution of the DF takes the form
\[
{\pa f\over\pa t}=-{\pa\over\pa\vJ}\cdot\vF,
\]
 where $\vF$ is the diffusive flux of stars in action space.  If $\dot
P(\vJ,\vDelta)$ is the rate of increase with time of the probability that a
star  scatters from $\vJ$ to $\vJ+\vDelta$, then $\vF$ is given by
\citep{Binney1988}
 \[
F_i=f\overline{\Delta_i}-\fracj12{\pa f\overline{\Delta_{ij}^2}\over\pa J_j},
\]
 where the first- and second-order diffusion coefficients are
\[
\begin{aligned}
\overline{\Delta_i}(\vJ)&=\int\d^3\vDelta\,\Delta_i\dot P(\vJ,\vDelta)
\\
\overline{\Delta_{ij}^2}(\vJ)&=\int\d^3\vDelta\,\Delta_i\Delta_j\dot
P(\vJ,\vDelta).
\label{first_second_diffusion_coeff}
\end{aligned}
\]
\cite{Binney1988} showed that in the relevant circumstances the first- and
second-order
diffusion coefficients are related by
 \[
\overline{\Delta_i}=\fracj12{\pa\overline{\Delta_{ij}^2}\over\pa J_j},
\]
 so the diffusive flux can be written entirely in terms of the second-order
 coefficients:
\[
F_i=-\fracj12\overline{\Delta_{ij}^2}{\pa f\over\pa J_j}.
\label{definition_flux_total}
\] 

By expanding $\psi(\vx,t)$, the fluctuating part of the gravitational
potential, in
angle-action variables,
\[
\psi(\vx,t)=\psi(\vtheta,\vJ,t)=\sum_\vm\psi_\vm(\vJ,t)\e^{\i\vm\cdot\vtheta},
\]
 \cite{Binney1988} showed that the second-order diffusion coefficients are
 related to the fluctuations in the potential by
\[
\overline{\Delta_{ij}^2}(\vJ)=\sum_{\vm}m_im_j\widetilde
c_\vm(\vJ,\vm\cdot\vOmega(\vJ)),
\label{expression_diffusion_coefficients_all}
\]
 where an overline indicates an ensemble average and $\widetilde c_\vm$ is
the Fourier transform with respect to time of the autocorrelation of the
$\vm$ Fourier component of the potential
\[
\widetilde c_\vm(\vJ,\nu)=\int_{-\infty}^\infty
\d\tau\,\e^{\i\nu\tau}\,\overline{\psi_\vm(\vJ,t)\psi^*_\vm(\vJ,t-\tau)}.
\]
 By the Wiener--Khinchin theorem, $\widetilde c_\vm(\vJ,\nu)$ is the power spectrum of
the stationary random variable $\psi_\vm(\vJ,t)$.  Equation
(\ref{expression_diffusion_coefficients_all}) tells us that diffusion is
driven by resonances because it implies that the rate at which stars diffuse
from action $\vJ$ is proportional to the power that the fluctuating field has
at any of the orbit's characteristic frequencies $\vm\cdot\vOmega(\vJ)$.
Hence, if the fluctuations are confined to a narrow frequency range, perhaps
because they are associated with spiral arms, stars that respond strongly to
them will be located at only a few points in action space.

\subsection{A basis-function expansion}

We will find it expedient to expand the fluctuating potential $\psi(\vx,t)$
in a set of basis functions, the members of which are enumerated by an index
$q$:
\[
\psi(\vx,t)=\sum_qb_q(t)\si q(\vx),
\]
 where $b_q(t)$ is a random variable.
Following \cite{Kalnajs}, we require our basis
potentials to be orthonormal to the densities $\ho p(\vx)$ that generate them, so
we have
 \[\label{orthog}
\nabla^2\si p=4\pi G\ho q\hbox{ and }
\int\d^3\vx\,\ho p(\vx)[\si q(\vx)]^*=-\delta_{pq}.
\]
Now we have
\[
\begin{split}
\psi_\vm(\vJ,t)&={1\over(2\pi)^3}
\int\d^3\vtheta\,\e^{-\i\vm\cdot\vtheta}\psi(\vtheta,\vJ,t)\\
&=\sum_p b_p(t)\si p_\vm(\vJ),
\end{split}
\]
 where
\[
\si p_\vm(\vJ)\equiv {1\over(2\pi)^3}
\int\d^3\vtheta\,\e^{-\i\vm\cdot\vtheta}\si p[\vx(\vtheta,\vJ)].
\]
 Hence the required power spectrum is
\[\label{required_c}
\widetilde c_\vm(\vJ,\nu)=\sum_{pq} B_{pq}(\nu)\si p_\vm(\vJ){\si
q_\vm}^*(\vJ),
\]
 where
\[\label{eq:defsB}
B_{pq}(\nu)\equiv\int_{-\infty}^\infty
\d\tau\,\e^{\i\nu\tau}\,\overline{b_p(t)b_q^*(t-\tau)}
\] 
 is the Fourier transform of the cross-correlation of the amplitudes of the
$p$ and $q$ basis functions.

Below we shall require an expression for $B_{pq}(\nu)$ in terms of the
Fourier transform
\[
\widetilde b_p(\nu)=\int\d t\,\e^{\i\nu t}b_p(t)
\]
 of $b_p(t)$. If $\psi(t)$ is a stationary random process, then it is 
straightforward to show that
\[\label{cross_auto}
\overline{\widetilde b_p(\nu)\widetilde
b_q^*(\nu')}=2\pi\delta(\nu-\nu')B_{pq}(\nu).
\] 

\subsection{Bare and dressed stimuli}

As we indicated in the Introduction, a stellar disc is exposed to several
sources of fluctuations. The issue that we now have to confront is that the
fluctuation $\psi$ in the potential that stars experience, which is what
appears in the above formulae, differs from the original stimulation,
$\psi^\e$, because the disc has non-negligible mass, so through Poisson's
equation it makes a
contribution $\psi^\rs$ to the actual gravitational potential $\psi$~\citep{Weinberg2001a}.
We shall refer to $\psi^\e$ as the ``bare'' stimulus and to
\[\label{eq:dressed_bare}
\psi(t)=\psi^\e(t)+\psi^\rs(t)
\]
as the ``dressed'' stimulus.  We now seek a relationship between the dressed
and bare stimuli.

Let $\psi'$ be the change in the
potential that the disc would  generate if its particles moved in
the sum of the unperturbed potential and the stimulating potential $\psi^\e$.
Then $\psi'(\vx)$ is linearly related to $\psi^\e(\vx)$ so for each
time-lapse $\tau$ there is a
linear \textit{response operator} $M(\tau)$ that connects these functions
 \[\label{eq:undress}
\psi'(t)=\int_{-\infty}^t\d t'\,M(t-t')\psi^\e(t').
\]
 Since the mass of the disc actually contributes to the potential in which
its particles move, changes in the disc's potential at a early time $t'$
contribute alongside $\psi^\e(t')$ to the disturbance of disc particles at
later times, so the fluctuating component of the disc's potential $\psi^\rs$
satisfies
\[
\psi^\rs(t)=\int_{-\infty}^t\d t'\,M(t-t')[\psi^\rs(t')+\psi^\e(t')].
\]
 Inserting this expression into equation \eqref{eq:dressed_bare},
we obtain
\[\label{eq:dress}
\psi(t)=\psi^\e(t)+\int_{-\infty}^t\d
t'\,M(t-t')\psi(t').
\]
 In this equation the potentials are functions of $\vx$ as well as $t$ and
$M(t-t')$ is an operator that maps one function of space onto another. The
basis functions introduced above
reduce this operator to a matrix, so when we write
\[
\begin{split}
\psi^\e(\vx,t)&=\sum_p a_p(t)\si p(\vx)\\
\psi(\vx,t)&=\sum_p b_p(t)\si p(\vx),
\end{split}
\]
 equation (\ref{eq:dress}) can be written
\[\label{eq:undress2}
\begin{split}
\sum_pb_p(t)\si p(\vx)=&\sum_q\Big[a_q(t)\si q(\vx)\\
&+\int_{-\infty}^t\d t'M(t-t')b_q(t')\si q(\vx)\Big].
\end{split}
\]
We multiply both sides of this equation by $-\int\d^3\vx \, [ \ho
r ]^{*}$ and with equation (\ref{orthog}) obtain
\[\label{dress2}
b_r(t)=a_r(t)+\sum_q\int_{-\infty}^t\d t'M_{rq}(t-t')b_q(t'),
\]
 where
\[
M_{rq}(t-t')=-\int\d^3\vx\, [ \ho r(\vx) ]^{*} \,M(t-t')\, \si q(\vx).
\]
 The temporal convolution can be reduced to a multiplication by taking a
Fourier transform: multiplication of equation (\ref{dress2}) by $\int\d
t\,\e^{\i\nu t}$ yields 
 \[
\widetilde b_r(\nu)=\widetilde a_r(\nu)+\sum_q\widetilde
M_{rq}(\nu)\widetilde b_q(\nu).
\]
 Hence 
\[\label{dress_final}
\widetilde\vb(\nu)=[\vI-\widetilde\vM(\nu)]^{-1}\widetilde\va(\nu),
\]
 where boldface implies vectors and matrices indexed with $p$.

Equations (\ref{cross_auto}) and (\ref{dress_final}) enable us to relate
$B_{pq}$ to the basis coefficients of the stimulating field
\[
\vB(\nu)={1\over2\pi}\int\d\nu'\,[\vI-\widetilde\vM(\nu)]^{-1}\overline{\widetilde\va(\nu)\otimes\widetilde\va^*(\nu')}
[\vI-\widetilde\vM^\dagger(\nu')]^{-1}.
\]
 We will show below that analogously to equation (\ref{cross_auto})
\[\label{def_bigA}
\overline{\widetilde a_p(\nu)\widetilde a_q^*(\nu')}
=2\pi\delta(\nu-\nu')A_{pq}(\nu).
\]
Hence equation (\ref{required_c}) can be written
\[\label{key_cm_to_a}
\begin{split}
\widetilde c_\vm(\vJ,\nu)
=&\sum_{pq}\si p_\vm(\vJ){\si
q_\vm}^*(\vJ)\\
&\times\left\{[\vI-\widetilde\vM(\nu)]^{-1}\vA(\nu)
[\vI-\widetilde\vM^\dagger(\nu)]^{-1}\right\}_{pq}.
\end{split}
\]
Our derivation of the dressed secular diffusion coefficients
sketched previously is based on the master equation~\citep{BT08}
which led to the first- and second-order diffusion coefficients from equation~\eqref{first_second_diffusion_coeff}.
One can also recover these diffusion coefficients via a timescale decoupling
of the collisionless Boltzmann equation~\citep{Weinberg2001a,Pichon2006,Chavanis2012EPJ,FouvryPichonPrunet2014}.
Various sources of external perturbations can then be considered
to induce secular evolution~\citep{Weinberg2001b,AubertPichon2007}.

\section{Application to a cool, thin disc}
\label{sec:appl}

The simplest non-trivial context in which the above principles can be
illustrated is the case of a cool razor-thin disc, i.e., a disc in
which every star is confined to a plane by $J_z=0$ and orbits have only
moderate eccentricities.  In this case each unperturbed orbit is
characterised by two numbers $(J_r,J_\phi)$ or a point $\vJ$ in
two-dimensional action space.  The angular momentum $J_\phi$ is as ever a
trivial function of $(\vx,\vv)$, and since orbits have only moderate
eccentricities, the epicycle approximation provides an adequate expression
for $J_r(\vx,\vv)$. If $\kappa(J_\phi)$ denotes the radial epicycle
frequency, the disc's unperturbed DF can be taken to be an exponential
$\exp(-\kappa J_r/\sigma_r^2)$ in $J_r$ times a function of $J_\phi$ that
essentially controls the disc's radial surface-density profile.  Then within
the epicycle approximation the velocity distribution at any point in the disc
is a biaxial Gaussian with radial dispersion $\sigma_r$.  Because gas tends
to flow on nearly closed orbits, stars are born on
nearly circular orbits, i.e., along the $J_\phi$ axis of action space, and
diffuse from there in to the body of action space. We shall show that on
account of resonances, this diffusion can form ridges in action space.

\subsection{Choice of the basis}

Since we are working in two dimensions, the basis functions $\si p(\vx)$
become functions $\si p(R,\phi)$ of plane polar coordinates that are
orthonormal to the generating surface densities $\Igma p(R,\phi)$. Our
problem is simplified if we can choose the basis $\si p$ such that the
response operator $\widetilde M$ is diagonal and we now show that this is possible. 

It is well known that the potential generated via Poisson's equation by a
tightly wound spiral wave is itself a spiral wave with the same
wavevector $\vk=(k_r,k_\phi)$ \citep[e.g.][\S6.2.2]{BT08}. Moreover, the computation of the
dynamical response of  particles within our disc to a
tightly-wound perturbing potential that oscillates at angular frequency
$\nu$ is covered by standard texts (see, for example, \citealt{BT08}
\S6.2.2(d)
or \citealt{Tenerife} \S4.2 for two different approaches). The result is that
a spiral potential $\si p(\vx)\e^{\i\nu t}$ creates a spiral perturbation in the
surface density of test particles, which through Poisson's equation creates a
response potential $\psi'(\vx)\e^{\i\nu t}$ that differs from the original
stimulating potential only in magnitude. In fact
\[\label{proportion}
\psi'(R,\phi)=\lambda_\vk \si p(R,\phi),
\]
 where
\[\label{expression_eigenvalues}
\lambda_\vk= \frac{2 \pi G \Sigma |k_{r}|}{\kappa^{2} (1 - s^{2})}
\mathcal{F} (s,\chi) \,.
\]
 Here $\Sigma$ is the disc's surface density,
\[
s \equiv \frac{\nu - k_{\phi} \, \Omega_{\phi}}{\kappa}
\label{definition_s}
\]
 is the ratio of the frequency at which a star experiences the perturbation to
the epicycle frequency,
and $\mathcal{F}$ is the reduction factor  \citep{Kalnajs1965,Lin1966}
\[
\mathcal{F} (s,\chi) \equiv 2 \, (1 - s^{2}) \frac{\e^{- \chi}}{\chi}
\sum\limits_{m_r = 1}^{+ \infty} 
\frac{\mathcal{I}_{m_r} (\chi)}{ 1 - s^2/m_r^2} \, ,
\label{definition_fonction_F}
\]
 where $\mathcal{I}_{m_{r}}$ is a modified Bessel function and the dimensionless quantity
\[
\chi \equiv \frac{\sigma_{r}^{2} \, k_{r}^{2}}{\kappa^{2}}
\label{definition_chi}
\]
 is a measure of how warm the disc is. In cases of interest the reduction
factor is a number slightly smaller than unity and of little interest.

The proportionality (\ref{proportion}) suggests that in a basis formed of
tightly-wound spiral waves the Fourier transformed response operator
$\widetilde M(\nu)$ is diagonal with $\lambda_\vk$ the diagonal element
associated with the given spiral wave. Hence the natural procedure might seem
to be the adoption of the complete set of logarithmic spirals
\citep[e.g.][\S2.6.3]{BT08} as the basis $\si p$.  Unfortunately, $\widetilde
M(\nu)$ is not, in fact, diagonal in the basis formed by logarithmic
spirals for the following reason. The demonstration that a spiral
perturbation generates a spiral response scaled by $\lambda_\vk$ is a {\it local}
result: the disc is analysed in just an annulus, and in the spirit of
WKB analysis, the wave considered is a packet of finite length. Since the
frequencies $\kappa$ and $\Omega_\phi$ that appear in equation
(\ref{definition_s}) are functions of radius, $\lambda_\vk$ is also a function of
radius whereas a diagonal element of $\widetilde M(\nu)$ should be a
constant. Hence only a short packet of spiral waves provides a good
approximation to an eigenfunction of $\widetilde M(\nu)$. Such a packet is
a non-trivial superposition of logarithmic spirals, so $\widetilde M(\nu)$
cannot be diagonal in the basis provided by logarithmic spirals. Physically,
the dynamics of the disc is inherently local on account of the existence of
resonant radii, so basis functions such as logarithmic spirals that extend
from the disc's centre to infinity cannot make $\widetilde M(\nu)$ diagonal.
Given that we want $\widetilde M(\nu)$ to become diagonal, we must work with
basis functions $\si p$ that are local.

\cite{FouvryPichonPrunet2014} show how to construct a biothorgonal basis of localised
spirals. They divide the range $(R_{\rm min},R_{\rm max})$ of relevant radii
into intervals of width $\sigma$ centred on $R_0$, and then for any given
wavevector $\vk=(k_r,k_\phi)$ create a basis function for each interval.
Specifically, their basis potentials are
\[
\psi^{(\vk,R_0)}(R,\phi)=\sqrt{{G\over|k_r|R_0}}
{\e^{\i(k_rR+k_\phi\phi)}\over(\pi\sigma^2)^{1\!/\!4}}
\exp\left[-{(R-R_0)^2\over2\sigma^2}\right] \, ,
\label{definition_basis_WKB}
\]
 with $k_rR_0\gg1$. The corresponding surface densities are
\[
\Sigma^{(\vk,R_0)}=-{|k_r|\over2\pi G}\psi^{(\vk,R_0)}.
\]
They show that two basis functions $\si{\vk^1,R_0^1}$ and
$\si{\vk^2,R_0^2}$ will be biorthogonal only when ${\Delta R_{0}
\equiv R_{0}^{1} - R_{0}^{2}}$ and ${\Delta k_{r} = k_{r}^{1} - k_{r}^{2}}$
satisfy
\[
\begin{cases}
\begin{aligned}
\displaystyle \Delta R_{0} &\gg \sigma\,, \;&\text{or}&\;\;\; \Delta R_{0} = 0 \,,
\\
\displaystyle \Delta k_{r} &\gg \frac{1}{\sigma}\,, \;&\text{or}&\;\;\; \Delta k_{r} = 0\,.
\end{aligned}
\end{cases}
\label{separation_assumption}
\]
That is, the centres of neighbouring bands have to be separated by more than
the width of a band, and within any band, adjacent wavenumbers have to differ
by enough to give a significant phase difference across the band.

Now that the basis potentials have been chosen, one can use the mapping
$(\vx,\vv)\mapsto(\vtheta,\vJ)$ provided by the epicycle approximation
\citep[e.g.][eq.~82]{Tenerife} to compute their Fourier
transforms with respect to the angle variables:
\[\label{Fourier_basis_WKB}
\begin{aligned}
\psi^{(\vk,R_0)}_{\bm{m}} (\bm{J}) 
=& \; \delta^{k_{\phi}}_{m_{\phi}} \, \e^{\i m_{r} \theta^{0}_{R}} \, 
\sqrt{{G\over|k_r|R_0}}\frac{1}{(\pi \sigma^{2})^{1\!/\!4}} \, \e^{\i k_{r} \Rg }
\\
& \times\mathcal{J}_{m_r} \left( \sqrt{\tfrac{\displaystyle 2
J_r}{\displaystyle \kappa}}k_{r}\right) \, \exp \left[ - \frac{(R_g - R_{0})^{2}}{2 \sigma^2} \right] \, ,
\end{aligned}
\]
 where $\Rg(J_\phi)$ is the radius of the circular orbit with angular
momentum $J_\phi$ and $\mathcal{J}_{m_{r}}$ is a Bessel function of the first
kind. On account of the tight-winding condition $k_{r} \Rg \gg 1$, the phase
shift $\theta^{0}_{R}$ is given by
\[
 \theta^{0}_{R} \simeq - \pi / 2 \, .
\label{expression_H_theta_0}
\]

In this basis, the response matrix $\widetilde M$ is diagonal, having diagonal
elements
\[\label{expression_M_diagonal_tepid}
\widetilde M_{\left(\vk^1,R_0^1 \right)
\left(\vk^2,R_0^2 \right)} 
= \delta_{k^{1}_r}^{k^{2}_r} \, \delta_{k^{1}_{\phi}}^{k^{2}_{\phi}} \,
\delta_{R_0^1}^{R_0^2}\,\lambda_\vk,
\]
 where $\lambda_\vk$ is defined by equation (\ref{expression_eigenvalues}).
This expression for $\widetilde\vM$ allows us to rewrite equation
(\ref{key_cm_to_a}) in the form
\[\label{new_key_cm_to_a}
\begin{split}
\widetilde c_\vm(\vJ,\nu)
=&\sum_{pq}\si p_\vm(\vJ){\si
q_\vm}^*(\vJ){A_{pq}(\nu)\over(
1-\lambda_{\vk^p})(1-\lambda_{\vk^q})}.
\end{split}
\]

In the chosen basis the expansion coefficients of the stimulating field are
\[\label{a_p_almost}
\begin{split}
a_p(\nu)&=-\int\d^2\vx\bigl[\Igma p(\vx)\bigr]^*\widetilde\psi^\e(\vx,\nu)\\
&=\sqrt{{|k^p_r|\over GR^p_0}}{1\over(\pi\sigma^2)^{1\!/\!4}}\int\d
R\,R\\
&\quad\times\exp\left[-{(R-R_0^p)^2\over2\sigma^2}\right]\e^{-\i
Rk_r^p}\widetilde\psi^\e_{k_\phi^p}(R,\nu),
\end{split}
\]
 where
\[
\widetilde\psi^\e_{k_\phi^p}(R,\nu)\equiv{1\over2\pi}\int\d\phi\,\e^{-\i
k_\phi^p\phi}\widetilde\psi^\e(R,\phi,\nu)
\]
 is the  Fourier transform in azimuthal angle and time of the stimulating potential. We
further define the local radial Fourier transform of $\psi^\e$ within the segment
centred on $R_0^p$ by \citep{Gabor1946}
 \[\begin{split}
\widetilde\psi^\e_{\vk^p}(R_0^p,\nu)\equiv&{1\over2\pi}\!\!\int\!\!\d R\,
\exp\left[-{(R-R_0^p)^2\over2\sigma^2}\right]\\
&\times\e^{-\i(
R-R_0^p)k_r^p}\widetilde\psi^\e_{k_\phi^p}(R,\nu).
\end{split}\]
 This definition is motivated by the consequence that thus defined the local radial
Fourier transform of a uniform potential $\psi^\e=1$ is independent of $R_0^p$.
If in equation (\ref{a_p_almost}) we approximate the leading factor $R$ in
the integrand by $R_0$, we then have
 \[
\widetilde a_p(\nu)=\sqrt{{|k_r^p|R_0^p\over
G}}{2\pi\over(\pi\sigma^2)^{1\!/\!4}}\e^{-\i
R_0^pk_r^p}\widetilde\psi^\e_{\vk^p}(R_0^p,\nu).
\]
 We require the ensemble average  $\overline{\widetilde a_p(\nu)\widetilde
 a_q^*(\nu')}$
(eqs.~\ref{def_bigA} and \ref{key_cm_to_a}), which is related to the ensemble average
$\overline{\psi^\e(\vx,t)\psi^\e(\vx',t')}$. We assume that stimulating
fluctuations are quasi-stationary in the sense that
 \[
\overline{\psi^\e_{k_\phi}(R,t)\psi^{\e*}_{k_\phi}(R',t')}=
C_{k_{\phi}} (t-t',R-R',(R+R')/2),
\]
 with the dependence on $R+R'$ being weak. With this assumption that the
process
$\psi^\e_{k_\phi}$ is stationary in time and ``locally stationary'' in space,
it follows that
\[
\begin{split}
\overline{\widetilde\psi^\e_{\vk^p}(R,\nu)\widetilde\psi^{\e*}_{\vk^q}(R',\nu')}&
=2\pi\delta(\nu-\nu')\delta(k_r^p-k_r^q)\\
&\times C(\vk^p,\nu,(R+R')/2),
\end{split}
\]
 where $C(\vk,\nu,R)$ is the spatio-temporal power spectrum of the
stimulating noise in the neighbourhood of $R$. Now we can write
 \[\label{eq:oscil}
\begin{split}
\overline{\widetilde a_p(\nu)\widetilde a_q^*(\nu')}&={|k_r^p|R_0^p\over G}
{(2\pi)^3\over(\pi\sigma^2)^{1\!/\!2}}
\e^{-\i k_r^p(R^p_0-R^q_0)}
\delta(\nu-\nu')\\
&\times\delta(k_r^p-k_r^q)C(\vk^p,\nu,(R_0^p+R_0^q)/2),
\end{split}
\]
 so by equation (\ref{def_bigA})
\[
\begin{split}
A_{pq}(\nu)&={|k_r^p|R_0^p\over G}
{(2\pi)^2\over(\pi\sigma^2)^{1/2}}
\e^{-\i k_r^p(R^p_0-R^q_0)}\\
&\times\delta(k_r^p-k_r^q)C(\vk^p,\nu,(R_0^p+R_0^q)/2).
\end{split}
\]

To obtain the diffusion coefficients from
equations~\eqref{expression_diffusion_coefficients_all} we substitute into
equation (\ref{new_key_cm_to_a}) for $\widetilde c_\vm$ our expressions
(\ref{Fourier_basis_WKB}) for $\si p_\vm(\vJ)$ and the above
expression for $\vA$. 
We have
\[\label{there}
\begin{split}
\widetilde c_\vm(\vJ,\nu)
=&\frac{(2\pi)^2}{\pi\sigma^{2}}\sum_{pq}
\delta^{k_\phi^p}_{m_\phi}\delta^{k_\phi^q}_{m_\phi}
\mathcal{J}_{m_r}^2\!\!\left(\!\sqrt{\frac{2J_r}{\kappa}}k_r^p\!\right)
\e^{-\i k_r^p(R^p_0-R^q_0)}\\
& \times
\exp\left[ - \frac{(\Rg -R_0^p)^2+(\Rg -R_0^q)^2}{2 \sigma^2} \right]\\
&\times\delta(k_r^p-k_r^q){C(\vk^p,\nu,(R_0^p+R_0^q)/2)\over(
1-\lambda_{\vk^p})^2}.
\end{split}
\]
 The sums over $p$ and $q$ expand into sums over $\vk^p$, $\vk^q$, $R_0^p$
and $R_0^q$. On account of the factors $\delta^{k_\phi^p}_{m_\phi}$ and
$\delta^{k_\phi^q}_{m_\phi}$ the sums over the $\phi$ components of the
$\vk^i$ are trivial. The sums over the $k_r^i$ and $R_0^i$ we approximate
with integrals by the substitutions
\[
\begin{split}
\sum_{k_r} f(k_r) &\rightarrow {1\over\Delta k_r}\int\d k_r\, f(k_r),\\
\sum_{R_0} g(R_0) &\rightarrow {1\over\Delta R_0}\int\d R_0\, g(R_0),
\end{split}
\]
 where $\Delta k_r$ is the difference between successive values of $k_r$ in
 the sum, and similarly for $\Delta R_0$. Then the Dirac delta function in
equation (\ref{there}) allows us to integrate over $k_r^q$. We assume that
$\sigma$ is small enough that we can approximate each Gaussian exponential by
$\sqrt{2\pi}\sigma\delta(\Rg-R_0^i)$ so we can trivially integrate over the
$R_0^i$. Finally, the presence in equation (\ref{eq:oscil}) of a rapidly
oscillating complex exponential $\e^{\i k_rR_0}$ imposes that the intervals
$\Delta k_r$ and $\Delta R_0$ must satisfy the critical-sampling condition
$\Delta k_r^i\Delta R_0^i=2\pi$ \citep{Daubechies1990}.  With this condition,
equation (\ref{there}) simplifies to
 \[\label{WKB_c_tilde}
\widetilde c_\vm(\vJ,\nu)=2\int\d k_r\,
{\mathcal{J}^2_{m_r}\!\left(\!\sqrt{2J_r/\kappa}\,k_r\!\right)
\over(1-\lambda_\vk)^2}\,
C(\vk,\nu,\Rg),
\]
 where by hypothesis the
dependence of $C$ on $\Rg$ is weak, and we have ${ k_{\phi} = m_{\phi} }$. This is our principal result. It enables us
to compute the diffusion tensor at any point in the action space of a thin,
self-gravitating disc given the power spectrum of the noise that excites
spiral structure in the disc.

\section{Application to a Mestel disc}
\label{sec:model}

We apply our results to the same  Mestel disc 
\citep{Mestel1963} that S12 discussed. The basic properties of this disc are
given in \S2.6.1(a) and \S4.5.1 of \cite{BT08}. Its circular speed is a
constant $V_0$ and its potential is
\[
\psi (R) = V_{0}^{2} \ln \!\left[ \frac{R}{R_{\rm in}} \right] \,,
\label{potential_Sellwood}
\]
 where the value of $R_{\rm in}$ is arbitrary. The corresponding surface
density is
\[
\Sigma (R) = \frac{V_{0}^{2}}{2 \pi G R} \,.
\label{surface_density_Sellwood}
\]
 \cite{Toomre1977} gives DF $f(E,J_\phi)$ that self-consistently generates
this disc. When we use the epicycle approximation to replace the energy $E$
by the radial action $J_r$, the  DF becomes 
\[\label{DF_Jr_Jphi}
f_{0} (J_{r},J_{\phi}) =  C\Theta \left( J_{\phi} \right)
\left( \frac{J_{\phi}}{R_{\rm in} V_{0}} \right)^{-1} \!\!
\exp \!\left( - \frac{\kappa (J_{\phi})}{\sigma_{r}^{2}} J_{r} \right),
\]
where 
\[\label{expression_Cprime}
C = \frac{V_{0}^{q + 2}/(2 \pi G R_{\rm in})}{2^{q/2} \sqrt{\pi} (q/2 - 1/2) ! \,
\sigma_{r}^{q+2}}
 \, \exp\bigl(\! - {V_{0}^{2}}/{2 \sigma_{r}^{2}} \bigr) \,,
\]
and 
\[
q = \frac{V_{0}^{2}}{\sigma_{r}^{2}} - 1 \, ,
\label{expression_q_Sellwood}
\]
and ${\Theta  \left( J_{\phi} \right) }$ is an Heaviside function removing retrograde stars.

Since the central singularity and infinite extent of the Mestel disc are
problematic, it is customary to modify the DF (\ref{DF_Jr_Jphi}) by
multiplying it by factors $T(J_\phi)$ that taper the stellar distribution at
very small and very large radii. These factors are
\begin{equation}
\begin{split}
 T_{\rm in} (J_{\phi}) &= \frac{J_{\phi}^{\nu}}{(R_{\rm in} V_{0})^{\nu} + J_{\phi}^{\nu}} \,,
\\
 T_{\rm out} (J_{\phi}) &= \left[ 1 + \left( \frac{J_{\phi}}{R_{\rm out}
 V_{0}} \right)^{\mu} \right]^{-1} \!\!,
\end{split}
\label{inner_outer_taper_Sellwood}
\end{equation}
where $\nu$ and $\mu$ control the sharpness of the two tapers.
$T_{\rm in}$ models the presence of a bulge by diminishing the DF inward
of $R_{\rm in}$.  Here, $T_{\rm out}$ models the outer edge of the disc, beyond which
the gravitational field is entirely generated by dark matter. Even after
tapering the stellar distribution, $\psi(R)$ continues to be given by
equation (\ref{potential_Sellwood}) because the bulge and the dark halo are
presumed to provide the gravitational force that was originally provided by
the un-tapered disc.

In our numerical work we use the same taper constants as S12. We adopt a
system of units such that: $V_0 = G = R_{\rm in} = 1$. The other numerical
factors are given by $q = 11.4$, $\nu = 4$, $\mu = 5$, $R_{\rm out} = 11.5$.

Within the epicyclic approximation, the azimuthal and 
radial frequencies are
\[
\begin{split}
 \Omega (J_{\phi}) &= \frac{V_{0}}{\Rg}\,,
\\
\displaystyle \kappa (J_{\phi}) &= \sqrt{2} \, \Omega (J_{\phi}) \,,
\end{split}
\label{expression_intrinsic_frequencies}
\]
 and are thus independent of $J_r$. The ratio ${\kappa}/{\Omega} = \sqrt{2}$
is a constant. This ratio determines the location of the resonances, so
it is important for the disc's dynamical behaviour. By taking it to be a
constant we risk introducing unphysical artifacts in the dynamics.

\begin{figure}
\begin{center}
\includegraphics[angle=-00,width=0.45\textwidth]{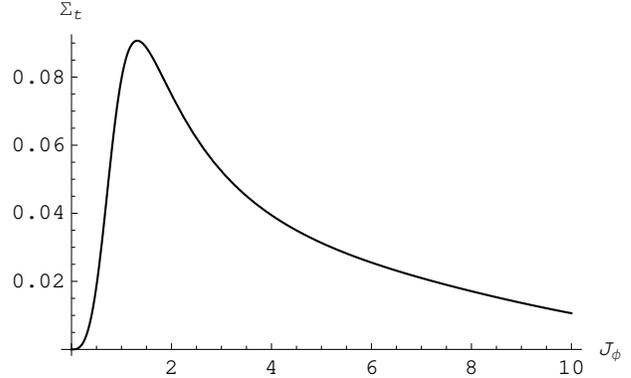}
\caption{\small{Surface density $\Sigma_{\rm t}$ of the tapered Mestel disc. The unit system has been chosen so that $V_{0} = G = R_{\rm in} = 1$.
}}
\label{figSigmaDF}
\end{center}
\end{figure}

\begin{figure}
\begin{center}
\includegraphics[angle=-00,width=0.45\textwidth]{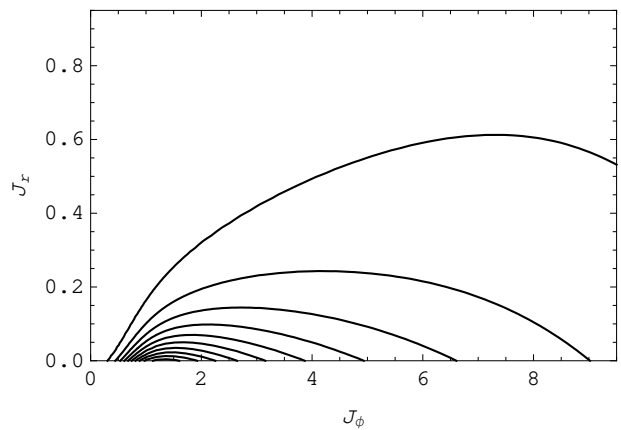}
\caption{\small{Contours of the initial distribution function in action-space $(J_{\phi},J_{r})$, within the epicyclic approximation. The contours are spaced linearly between 95\% and 5\% of the distribution function maximum.
}}
\label{figcontourDF}
\end{center}
\end{figure}

The distribution function~\eqref{DF_Jr_Jphi} multiplied by the taper factors
(\ref{inner_outer_taper_Sellwood}) takes the form of a locally
\textit{isothermal}-DF or Schwarzschild-DF with the correct normalization,
which can be rewritten as
\[
f_{0} (J_{r} , J_{\phi}) = \frac{\Omega (J_{\phi}) \, \Sigma_{\rm t}
(J_{\phi}) }{\pi \, \kappa (J_{\phi}) \, \sigma_{r}^{2}} \, \exp \left( -
\frac{\kappa (J_{\phi})}{\sigma_{r}^{2}} J_{r} \right) \,,
\label{DF_Jr_Jphi_isothermal}
\]
where the intrinsic frequencies are given by equation~\eqref{expression_intrinsic_frequencies} and the taped surface density in analogy with equation~\eqref{surface_density_Sellwood} is given by
\begin{equation}
\Sigma_{\rm t} (J_{\phi}) = \frac{V_{0}^{3}}{2 \pi G  J_{\phi}} \,
\Theta(J_\phi)\,T_{\rm in} (J_{\phi}) \, T_{\rm out} (J_{\phi}) \,.
\label{Sigma_taped}
\end{equation}
 The shape of the damped surface density is shown in Figure~\ref{figSigmaDF}.
 Figure~\ref{figcontourDF} shows the
level contours of the distribution function $f_0$.

\begin{figure}
\begin{center}
\includegraphics[angle=-00,width=0.45\textwidth]{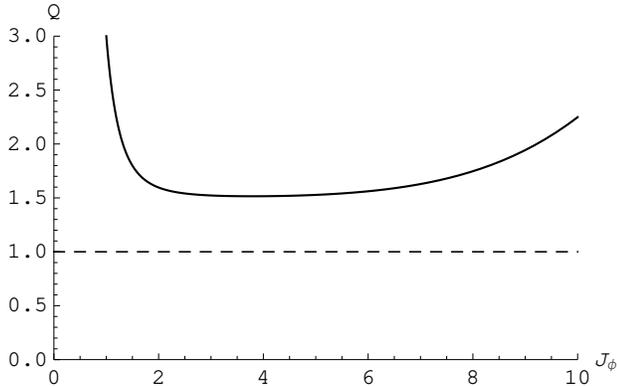}
\caption{\small{Variation of the Toomre parameter $Q$ with the angular
momentum $J_{\phi}$. It is scale invariant except in the inner/outer regions
because of the presence of the tapering functions $T_{\rm in}$ and $T_{\rm
out}$. The unit system has been chosen so that $V_{0} = G = R_{\rm in} = 1$.
}} \label{figQDF}
\end{center}
\end{figure}

In the scale-invariant Mestel disc the local \cite{Toomre1964} parameter
\[
Q = \frac{\sigma_{r} \, \kappa (J_{\phi})}{3.36 \, G \, \Sigma (J_{\phi})}
\label{definition_Q}
\]
is independent of radius, and in the tapered disc $Q$ is correspondingly flat
between the tapers. For realistically small values of $\sigma_r/V_0$, the
plateau in $Q$ can lie below unity, making the disc unstable. To keep $Q$
everywhere well above unity it is conventional to suppose that only a
fraction $\xi<1$ of the disc is self-gravitating with the rest of the
gravitational field provided by an unresponsive halo. In the S12 simulation,
the fraction of active surface density was $\xi = 0.5$. The dependence of $Q$
with radius with this value of $\xi$ is shown in Figure~\ref{figQDF} --
$Q\simeq1.5$ between the tapers and increases strongly in the tapered
regions.

\subsection{Impact of shot noise}

To proceed further we need to assume some form for the power spectrum
$C(\vk,\nu,\Rg)$ that appears in equation (\ref{WKB_c_tilde}). An inevitable
source of noise is shot noise caused by the the finite number of stars in the
disc, and massive, compact gas clouds are a source of spectrally similar
noise, so let us investigate the impact that shot noise has. In this case the
power spectrum is independent of $\nu$ and $k_r$, and varies with radius like
$\surd\Sigma(R)$. Then $C\propto\Sigma$, so to within a normalisation that
depends on particle number, we have
\[\label{diffusion_coefficients_WKB}
\widetilde c_\vm (\vJ,\nu) = \Sigma_{\rm t}(J_\phi)
\int \! \d k_{r} \, {\mathcal{J}^2_{m_r}
\!\left(\sqrt{2J_r/\kappa}\,k_r\right)\over(1 - \lambda_\vk)^2}.
\]
The eigenvalues $\lambda_\vk$ have to be evaluated at $\nu = \bm{m}
\!\cdot\! \bm{\Omega}$, and then $s = m_r$ by equation (\ref{definition_s}). In
order to handle the singularity of the
equation~\eqref{expression_eigenvalues} when $s = \pm 1$, one adds a small
imaginary part to the frequency of evaluation, so that $s = m_{r} + \i\eta$.
As long as $\eta$ in modulus is small compared to imaginary part of the least
damped mode of the disc, adding this complex part makes a negligible
contribution on the expression of $\text{Re}(\lambda)$. 

In equation \eqref{diffusion_coefficients_WKB} the integral over $k_r$ should
formally be over the full range $0$ to $\infty$. However, small values of
$k_r$ are unphysical and violate our assumption of tightly-wound spirals.
Values of $k_r$ that are larger than $\sim2\pi$ divided by the thickness of
a galactic disc are also unphysical, and in the case $m_r=0$ of the CR the
integral diverges at $J_r=0$ since then the Bessel function remains non-zero
to arbitrarily high $k_r$ and $(1-\lambda_\vk)^{-2}$ is always greater than
unity. Hence we must determine appropriate upper and lower limits to the
integration on $k_r$.

\begin{figure}
\begin{center}
\includegraphics[angle=-00,width=0.45\textwidth]{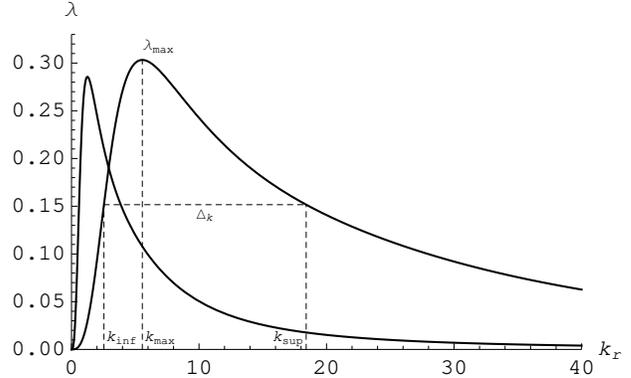}
\caption{\small{Variation of eigenvalues $\lambda$ of the response matrix with
the WKB-frequency $k_{r}$ for two values of $J_{\phi}$. The curve that peaks
at small $k_r$ is for the larger value of $J_\phi$.}} \label{figLambdakr}
\end{center}
\end{figure}

At any point in action space the biggest contribution to the diffusion tensor
will come from waves that yield the largest value of $\lambda_\vk$. Hence we
now examine the structure of the function $k_{r} \mapsto \lambda_\vk$ for
given $\vm$ and $\vJ$.  Figure~\ref{figLambdakr} shows that $\lambda_\vk$ has
a well-defined peak at a value $k_{\rm max}$ of $k_r$ that decreases as
$J_\phi$ increases. In fact $k_{\rm max}\propto1/J_\phi$ because the radius
of a near-circular orbit is $R\propto J_\phi$ and in a scale-free model we
expect $k_{\rm max}\propto1/R$. For the same reason we expect the width
$\Delta_k$ of the peak in $\lambda_\vk$ to be proportional to $1/J_\phi$.
In light of these observations we adopt as lower and upper bounds on the
$k_r$ integral the wavenumbers $k_{\rm inf}$ and $k_{\rm sup}$ defined by
\[
\lambda_{k_{\rm inf}}=\lambda_{k_{\rm sup}}= \fracj12\lambda_{k_{\rm
max}}.
\]
 These limiting values of $k_r$ are marked on Fig.~\ref{figLambdakr}.

\begin{figure}
\begin{center}
\includegraphics[angle=-00,width=0.45\textwidth]{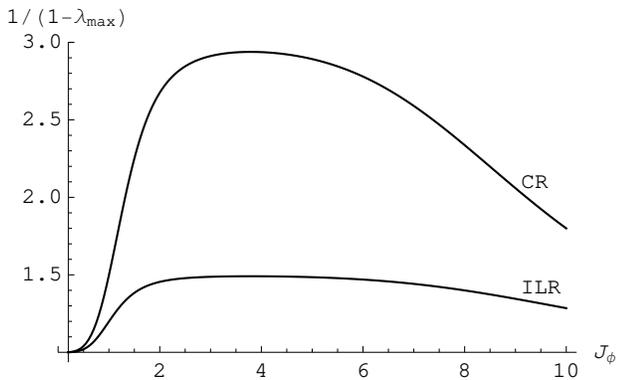}
\caption{\small{Dependence of the amplification factor $1/(1 - \lambda_{\rm
max})$ with the position $J_{\phi}$.  Throughout the disc, the
amplification of waves at corotation is larger than that of waves at inner
Lindblad resonance.}} \label{figLambdaJphi}
\end{center}
\end{figure}

At each point in action space there are contributions to the diffusion
coefficients from several values of $\vm$. The contribution from $m_r = -1$
is driven by waves that have their inner Lindblad resonance at that point,
while that from $m_r=+1$ is driven by waves that have their outer Lindblad
resonance there, and the contribution from $m_r=0$ is driven by waves locally
in corotation.  Since $\lambda_\vk$ depends on $s^2=m_r^2$, the
value of $\lambda_\vk$ is the same for $m_r=\pm1$. At a given point in
action space different values of the frequency $\nu$ are associated with
$m_r=\pm1$, but since we are considering shot noise, the fluctuations have the
same power at all frequencies. Hence the values $m_r=\pm1$ contribute equally to the
diffusion coefficients. The lower curve in Fig.~\ref{figLambdaJphi} shows
the extent to which this contribution is amplified by the disc's self-gravity
and we see that the amplification is modest. The upper curve in
Fig.~\ref{figLambdaJphi} shows the amplification by self-gravity in the
case $m_r=0$ of corotation: it is much larger.

In a cool disc, $|\pa f/\pa J_r|\gg|\pa f/\pa J_\phi|$ so by equation
(\ref{definition_flux_total}) a reasonable
approximation to the diffusive flux, when ${ m_{r} \neq 0 }$,
is 
\[
F_i=-\fracj12\sum_\vm m_i\widetilde c_\vm m_r{\pa f_0\over\pa J_r}.
\]
 It follows that waves at inner Lindblad resonance, which couple through
 $(m_r,m_\phi)=(-1,2)$ drive a diffusive flux
\[
\vF=-(-1,2)\fracj12\widetilde c_{(-1,2)}\left|{\pa f_0\over\pa J_r}\right|,
\]
 while waves at outer Lindblad resonance drive the flux
 \[
\vF=(1,2)\fracj12\widetilde c_{(1,2)}\left|{\pa f_0\over\pa J_r}\right|.
\]
 These formulae show that waves with a Lindblad resonance at $\vJ$ drive a
flux towards increasing $J_r$: these waves are heating the disc by increasing
the eccentricities of orbits. Wave that have their ILR at $\vJ$ drive stars
to lower angular momentum, while those with their OLR at $\vJ$ drive stars to
higher angular momentum. We shall see below that at low $J_\phi$ waves with a
local ILR dominate, so on average angular momenta decrease, while at large
$J_\phi$ the waves with a local OLR dominate and angular momenta increase on
average. Thus spiral structure conveys angular momentum outwards, thus
liberating energy that is used to heat the disc.

Waves at corotation couple through $\vm=(0,2)$ so (a) they can only drive
diffusion parallel to the $J_\phi$ axis, and (b) that diffusion is
proportional to the gradient's small component $\pa f/\pa J_\phi$.  In
practice this diffusion is important only in so far as the disc has a
metallicity gradient, when the DF of stars of any particular metallicity can
have a significant derivative with respect to $J_\phi$ and the large value of
$\widetilde c_\vm$ at corotation can have a big impact on the metallicity
distribution  -- radial migration at
corotation is important for chemical evolution
\citep{SellwoodB,Schonrich20091}.

\begin{figure}
\begin{center}
\includegraphics[angle=-00,width=0.90\columnwidth]{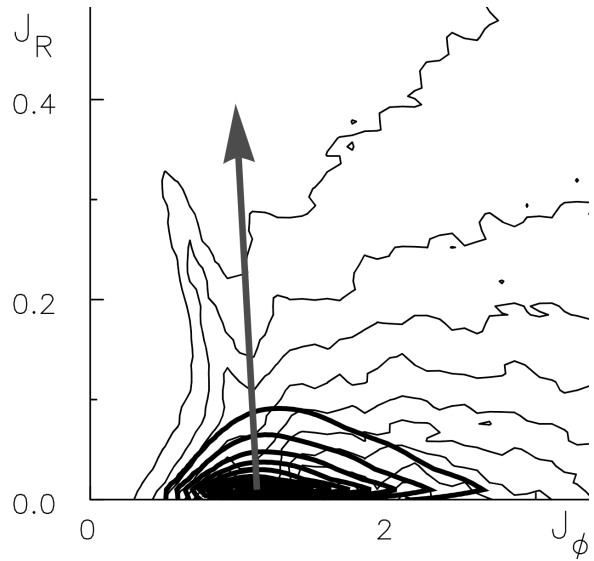}
 \caption{\small{Map of the norm of the total flux summed over the three
resonances (ILR, CR, OLR) (\textit{bold lines}). The contours are spaced
linearly between 95\% and 5\% of the function maximum. The gray vector gives
the direction of the vector flux associated to the norm maximum (arbitrary
length). The background contours correspond to the \textit{diffused}
distribution from S12 (\textit{thin lines}), which exhibits a narrow ridge of
diffusion.}} \label{figNorm_Flux}
\end{center}
\end{figure}
\subsection{Reproducing the S12 simulation}
\label{sec:S12results}

In this section we examine the extent to which our analytic results can
explain the simulations of tapered Mestel discs described by S12.  We do not
expect perfect agreement between our results and the numerical experiments
because we have employed several approximations. In particular we have
assumed that the driving fluctuations are white noise that has an amplitude
squared that is proportional to the disc's surface density. We have assumed,
moreover, that these fluctuations drive tightly-wound waves. S12 restricted
disturbing forces to $m_\phi=2$, so we impose this same restriction on $\vm$.

The dark contours in Fig.~\ref{figNorm_Flux} show the magnitude of the
diffusive flux that is generated by the two Lindblad resonances and
corotation when one adopts the same numerical pre-factors as S12. The gray
arrow shows the direction of the diffusive flux at the location of the peak
flux; the direction is similar at neighbouring points.
 The thin contours in Fig.~\ref{figNorm_Flux} show the value of the
distribution function in S12 after diffusion has taken place. Originally the
DF peaked along the $J_\phi$ axis at $J_\phi\simeq1$, becoming small to the
left of that point on account of the inner taper. Above the location of the
peak DF, there is a prominent horn of enhanced stellar density that extends
roughly in the direction of the diffusive flux. Thus our analytic results are
in good qualitative agreement with the numerical experiments of S12.

In Fig.~\ref{figNorm_Flux} waves at ILR drive diffusion parallel to vectors
that are inclined by $153^\circ$ to the $J_\phi$ axis, whereas the net flux
makes an angle of $111^\circ$ with the axis. That these angles are similar
attests to the dominant role of waves that have ILR near where the DF peaks.
This dominance arises because $\pa f_0/\pa J_r$ is always negative and in
this region $\pa f_0/\pa J_\phi>0$, so $|\vm\cdot{\pa f_0/\pa\vJ}|$ is larger
for the waves at ILR, which have $m_r=-1$, than for the waves at OLR, while
$\widetilde c_\vm$ is the same for both values of $m_r$.

Given that we have assumed that the driving fluctuations are white noise, it
is perhaps surprising that the magnitude of the diffusive flux is as sharply
peaked in action space as the dark contours in   Fig.~\ref{figNorm_Flux} show
it to be. This localisation surely reflects the sharp tapering of the DF at
small radii. Just outside this taper the disc becomes significantly
self-gravitating and most effective at amplifying the stimulating white
noise. The amplified waves tend to have their ILRs further in, where the
taper is cutting into the disc.

\begin{figure*}
\begin{center}
\begin{tabular}{@{}cc@{}}
\includegraphics[angle=-00,width=0.450\textwidth]{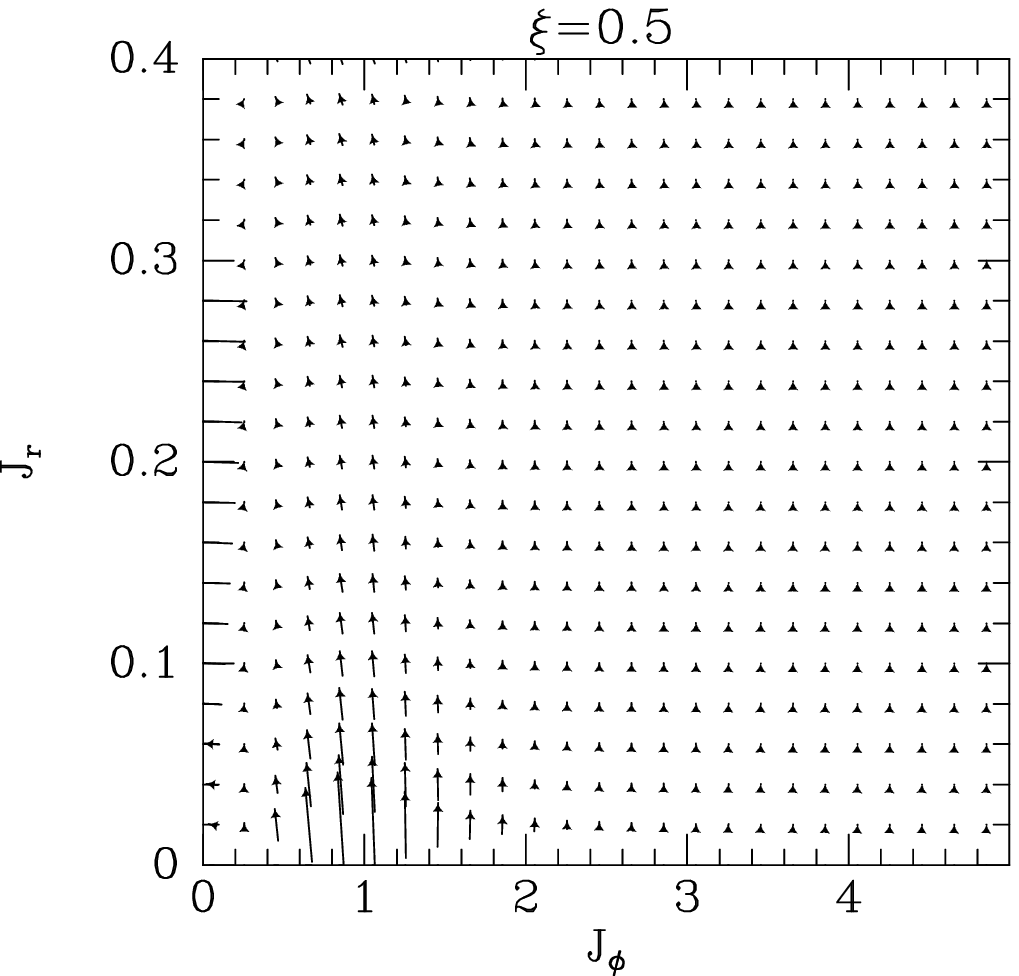} &
\includegraphics[angle=-00,width=0.450\textwidth]{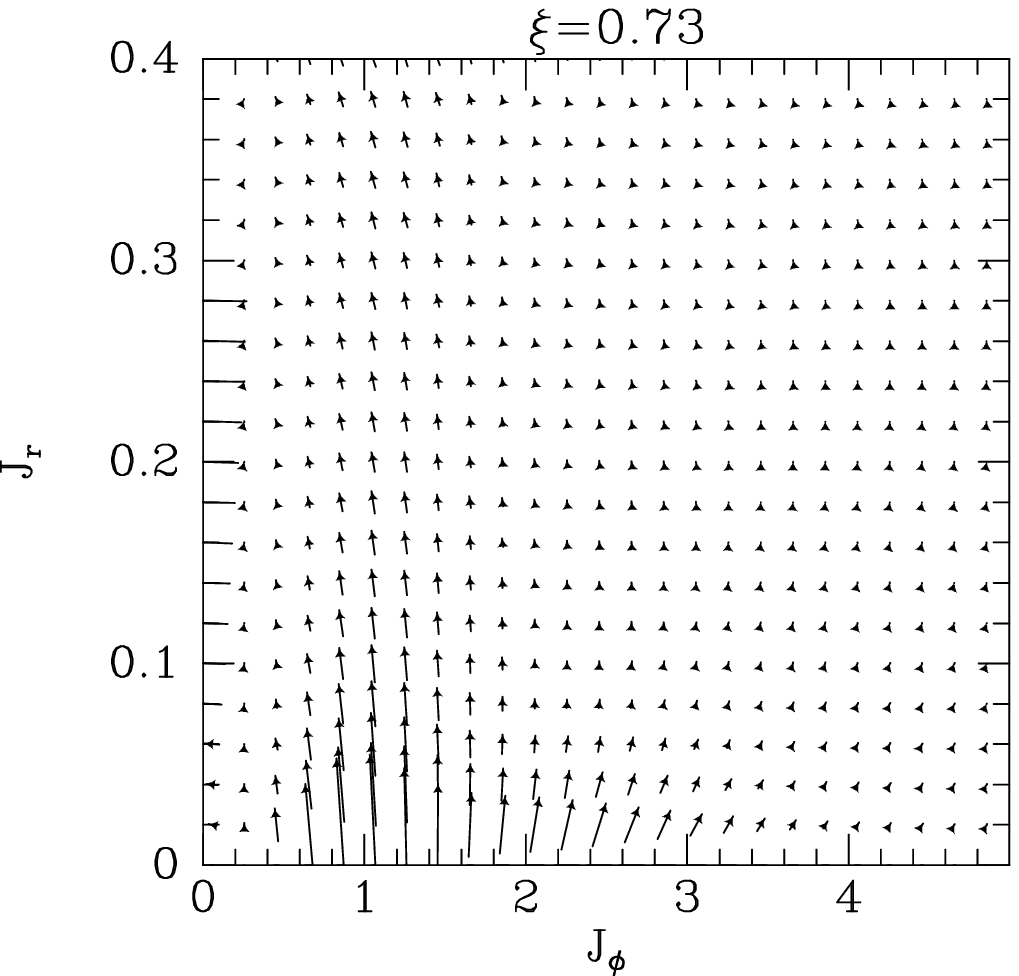}
\end{tabular}
 \caption{\small{Velocity $\vv\equiv \vF/f$ of the probability density in the $(J_\phi,J_r)$
 plane for two values of the active fraction of the disc mass: $\xi=0.5$
 (left) and $\xi=0.73$ (right). The flux vectors $\vF$ are obtained by summing 
over all three
resonances (ILR, CR, OLR).  As $\xi$ increases the CR becomes more important
and the main growth in $\vv$ occurs at $J_\phi\simeq2$.}} 
\label{figNorm_Flux_xi_all}
\end{center}
\end{figure*}

We have seen that self-gravity amplifies stimuli most strongly at corotation
(Fig.~\ref{figLambdaJphi}). In fact, the dominance of corotation increases
without limit as $Q$ decreases because the value of $\lambda_{\rm max}$
associated with corotation approaches unity, and the associated amplification
diverges, while the value of $\lambda_{\rm max}$ associated with the LRs
remains significantly below unity. Fig.~\ref{figNorm_Flux_xi_all} illustrates
this phenomenon by comparing the velocity of diffusion in action space,
\[
\vv={\vF\over f},
\]
 for two values of $\xi$. For the value $\xi=0.5$ assumed by S12 (left panel)
the velocity vectors are significantly non-zero only within the narrow strip
at $J_\phi\sim1$ associated with the ILR and point up and to the left. For
$\xi=0.73$ the velocity is also quite large at $J_\phi>2$ near the $J_\phi$
axis and is there directed up and to the right.

\section{Conclusions}
\label{sec:conclusions}

In a star cluster or a galaxy stars move on orbits in the system's mean
gravitational field, but the orbit each star is on evolves slowly in response
to fluctuations in the mean field \citep{Weinberg2001a}. In a hot stellar
system such as a star cluster, the dominant fluctuations are pure shot noise
arising from the finite number of stars in the system. In a disc galaxy the
situation is much more complex and interesting because the system's response
is more frequency-sensitive, and as Toomre's $Q\to1$  stimuli are
strongly amplified and distorted by the coherent responses they induce in the disc.

The orbit-averaged Fokker-Planck equation, which describes the evolution of
the distribution function as stars diffuse through action space, provides the
mathematical device of choice to compute the long-term evolution of a
stellar system. The equation is readily solved once the diffusion tensor has
been computed.

We have laid out the general formalism for computing the diffusion tensor in
the presence of significant coherent response to stimuli, and have
implemented this formalism in the case of a razor-thin, cool disc. By
introducing a set of basis functions for the disc's potential that comprise
sets of localised, tightly wound spirals, we have derived equation
(\ref{WKB_c_tilde}), which reduces computation of the diffusion tensor to
execution of a one-dimensional integral over the auto-correlation function of
the stimulating noise.

We used this equation to compute the diffusion tensor for a Mestel disc that
has been trimmed at both large and small radii by tapers and is exposed to
shot noise. We find that diffusive flux shows quite sharp peaks in action
space. When Toomre's parameter $Q$ is significantly bigger than unity, the
diffusive flux is quite localised near the inner taper, and is primarily
driven by waves that have their inner Lindblad resonances there. This result is
interesting in relation to the N-body simulations of S12 because in these
simulations, orbital diffusion led to the formation of a horn of enhanced
density in phase space that lies close to the region in which our analytic
work predicts that the diffusive flux peaks. Moreover, the direction
along the horn is similar to  the predicted  direction of the diffusive flux.
Thus it appears that our analytic work recovers quite well the main feature
of the simulations.

As $Q$ approaches unity, the corotation resonances of waves become more
important because self-gravity amplifies perturbations at corotation very
strongly. Waves that are in corotation at some point in action space drive
diffusion parallel to the angular-momentum axis of action space, i.e. they
drive radial migration. By contrast, waves that are at a Lindblad resonance
heat the disc by driving stars to higher eccentricity, while either
decreasing (at ILR) or increasing (at OLR) angular momentum. Consequently,
our calculations predict that the amount of radial migration for a given
level of heating increases as $Q$ decreases towards unity.

In our application of action-space diffusion to cool discs, the disc's
response to stimuli has been restricted to tightly-wound spirals. This
restriction unfortunately excludes from consideration the key physics of
``swing amplification'' at corotation \citep{Toomre1981}. As
\cite{GoldreichLB} originally showed in the context of a gas disc, leading
waves are amplified at corotation as they morph into trailing waves. These
waves will then propagate to the Lindblad resonances and there heat the disc
\citep{Toomre1981}.
Shot noise will stimulate leading waves in the same amount as trailing waves,
and the leading waves will move to corotation rather than to the Lindblad
resonances, and there morph into trailing waves of larger amplitude. Our
computation has not included the effect of swing-amplification at corotation,
and will consequently under-estimate the extent of heating. The
under-estimation will be largest for small values of $Q$, because swing
amplification diverges as $Q\to1$.

Another shortcoming of the present analysis is the crude noise model which
has ad-hoc dependence on the temporal frequency $\nu$ and the radial
wavenumber $k_r$, and is only a function of the position in the disc through
$J_{\phi}$. The model aims to reproduce the Poisson shot noise
caused by the finite number of particles in the disc. In a companion paper,
we will investigate the WKB limit of the Balescu-Lenard equation. This
approach will allow us to avoid such approximation since it naturally
captures the \textit{intrinsic noise}, due to finite$-N$ effects, and its
impact on the quasi-stationnary distribution function, as long as the
evolution of the system is made through transient tightly wound spirals.

In a forthcoming paper we will use the present formalism of forced diffusion to study the
evolution of a stellar disc as a result of cosmic noise: the noise generated by
satellites that orbit in a disc's host dark halo.

Another possible application of the formalism developed here is to study
the secular evolution of dark-matter cusps at the centres of galaxies in
response to stochastic excitation by the inner baryonic disc and bulge.
\citep{Fouvry2014}.

\vskip 0.5cm \noindent {\sl This paper is dedicated to
the memory of Jean Heyvaerts who was its  inspiration.}
\subsection*{Acknowledgements}

JBF thanks the {\sc GREAT} program for travel funding.  CP thanks the
Institute of Astronomy, Cambridge, for hospitality while this investigation
was initiated.  We thank S.~Colombi, S.~Prunet and P.~H.~Chavanis for
comments.  This work is partially supported by the Spin(e) grants
ANR-13-BS05-0005 of the French {\sl Agence Nationale de la Recherche} and by
the ILP LABEX (under reference ANR-10-LABX-63) which is funded by
ANR-11-IDEX-0004-02. The research
leading to these results has received funding from the European Research
Council under the European Union's Seventh Framework Programme
(FP7/2007-2013) / ERC grant agreement no.\ 321067.

\bibliographystyle{apj}
\bibliography{references}

\begin{thebibliography}{}
\expandafter\ifx\csname natexlab\endcsname\relax\def\natexlab#1{#1}\fi

\bibitem[{{Aubert} \& {Pichon}(2007)}]{AubertPichon2007}
{Aubert}, D., \& {Pichon}, C. 2007, \mnras, 374, 877

\bibitem[{{Aumer} \& {Binney}(2009)}]{AumerB}
{Aumer}, M., \& {Binney}, J.~J. 2009, \mnras, 397, 1286

\bibitem[{{Binney}(2010)}]{Binney10}
{Binney}, J. 2010, \mnras, 401, 2318

\bibitem[{{Binney}(2013)}]{Tenerife}
---. 2013, {Dynamics of secular evolution}, ed. J.~{Falc{\'o}n-Barroso} \&
  J.~H. {Knapen} (Cambridge University Press), 259

\bibitem[{{Binney}(2014)}]{Binney14}
---. 2014, \mnras, 440, 787

\bibitem[{{Binney} \& {Lacey}(1988)}]{Binney1988}
{Binney}, J., \& {Lacey}, C. 1988, \mnras, 230, 597

\bibitem[{Binney \& Tremaine(2008)}]{BT08}
Binney, J., \& Tremaine, S. 2008, Galactic Dynamics: (Second Edition),
  Princeton Series in Astrophysics (Princeton University Press)

\bibitem[{Born(1960)}]{born1960mechanics}
Born, M. 1960, The Mechanics of the Atom (F. Ungar Pub. Co.)

\bibitem[{{Chavanis}(2012)}]{Chavanis2012EPJ}
{Chavanis}, P.~H. 2012, European Physical Journal Plus, 127, 19

\bibitem[{Daubechies(1990)}]{Daubechies1990}
Daubechies, I. 1990, Information Theory, IEEE Transactions on, 36, 961

\bibitem[{{Dehnen}(1998)}]{Dehnen98}
{Dehnen}, W. 1998, \aj, 115, 2384

\bibitem[{{Eyre} \& {Binney}(2011)}]{EyreBinney}
{Eyre}, A., \& {Binney}, J. 2011, \mnras, 413, 1852

\bibitem[{{Famaey} {et~al.}(2005){Famaey}, {Jorissen}, {Luri}, {Mayor}, {Udry},
  {Dejonghe}, \& {Turon}}]{Famaey05}
{Famaey}, B., {Jorissen}, A., {Luri}, X., {et~al.} 2005, \aap, 430, 165

\bibitem[{{Fouvry} {et~al.}(2014{\natexlab{a}}){Fouvry}, {Pichon}, \&
  {Prunet}}]{FouvryPichonPrunet2014}
{Fouvry}, J.~B., {Pichon}, C., \& {Prunet}, S. 2014{\natexlab{a}}, in press

\bibitem[{{Fouvry} {et~al.}(2014{\natexlab{b}})}]{Fouvry2014}
{Fouvry}, J.~B., {et~al.} 2014{\natexlab{b}}, in prep

\bibitem[{Gabor(1946)}]{Gabor1946}
Gabor, D. 1946, Electrical Engineers - Part III: Radio and Communication
  Engineering, Journal of the Institution of, 93, 429

\bibitem[{{Goldreich} \& {Lynden-Bell}(1965)}]{GoldreichLB}
{Goldreich}, P., \& {Lynden-Bell}, D. 1965, \mnras, 130, 125

\bibitem[{{Helmi} \& {White}(1999)}]{HelmiWhite99}
{Helmi}, A., \& {White}, S.~D.~M. 1999, \mnras, 307, 495

\bibitem[{{Kalnajs}(1965)}]{Kalnajs1965}
{Kalnajs}, A.~J. 1965, Ph.D. thesis (Harvard University)

\bibitem[{{Kalnajs}(1971)}]{Kalnajs}
---. 1971, \apj, 166, 275

\bibitem[{{Lin} \& {Shu}(1966)}]{Lin1966}
{Lin}, C.~C., \& {Shu}, F.~H. 1966, Proceedings of the National Academy of
  Science, 55, 229

\bibitem[{{McMillan}(2013)}]{McMillan13}
{McMillan}, P.~J. 2013, \mnras, 430, 3276

\bibitem[{{Mestel}(1963)}]{Mestel1963}
{Mestel}, L. 1963, \mnras, 126, 553

\bibitem[{{Pichon} \& {Aubert}(2006)}]{Pichon2006}
{Pichon}, C., \& {Aubert}, D. 2006, \mnras, 368, 1657

\bibitem[{{Piffl} {et~al.}(2014){Piffl}, {Binney}, {McMillan}, {Bienaym{\'e}},
  {Bland-Hawthorn}, {Freeman}, {Gibson}, {Gilmore}, {Grebel}, {Helmi},
  {Kordopatis}, {Navarro}, {Parker}, {Reid}, {Seabroke}, {Siebert},
  {Steinmetz}, {Wyse}, \& {Zwitter}}]{Piffl14}
{Piffl}, T., {Binney}, J., {McMillan}, P.~J., {et~al.} 2014, ArXiv e-prints,
  arXiv:1406.4130

\bibitem[{{Sanders} \& {Binney}(2013)}]{SandersBinney}
{Sanders}, J.~L., \& {Binney}, J. 2013, \mnras, 433, 1826

\bibitem[{{Sch{\"o}nrich} \& {Binney}(2009)}]{Schonrich20091}
{Sch{\"o}nrich}, R., \& {Binney}, J. 2009, \mnras, 396, 203

\bibitem[{{Sellwood}(2010)}]{Sellwood10}
{Sellwood}, J.~A. 2010, \mnras, 409, 145

\bibitem[{{Sellwood}(2012)}]{Sellwood2012}
---. 2012, \apj, 751, 44

\bibitem[{{Sellwood} \& {Binney}(2002)}]{SellwoodB}
{Sellwood}, J.~A., \& {Binney}, J.~J. 2002, \mnras, 336, 785

\bibitem[{{Sellwood} \& {Carlberg}(2014)}]{SellwoodCarlberg14}
{Sellwood}, J.~A., \& {Carlberg}, R.~G. 2014, \apj, 785, 137

\bibitem[{{Toomre}(1964)}]{Toomre1964}
{Toomre}, A. 1964, \apj, 139, 1217

\bibitem[{{Toomre}(1977)}]{Toomre1977}
---. 1977, \araa, 15, 437

\bibitem[{{Toomre}(1981)}]{Toomre1981}
{Toomre}, A. 1981, in Structure and Evolution of Normal Galaxies, ed. S.~M.
  {Fall} \& D.~{Lynden-Bell}, 111--136

\bibitem[{{Weinberg}(2001{\natexlab{a}})}]{Weinberg2001a}
{Weinberg}, M.~D. 2001{\natexlab{a}}, \mnras, 328, 311

\bibitem[{{Weinberg}(2001{\natexlab{b}})}]{Weinberg2001b}
---. 2001{\natexlab{b}}, \mnras, 328, 321

\bibitem[{{Wielen}(1977)}]{Wielen77}
{Wielen}, R. 1977, \aap, 60, 263

\end{thebibliography}

\label{lastpage}
\end{document}